\documentclass[journal=jacsat,manuscript=article]{achemso}
\usepackage[fleqn]{amsmath}
\usepackage[version=3]{mhchem}
\title{Second harmonic generation in GaAs photonic crystal cavities in (111)B and (001) crystal orientations}

\author{Sonia Buckley}
\affiliation{Spilker Center for Engineering and Applied Sciences, Stanford
University, Stanford CA 94305}
\email{bucklesm@stanford.edu}
\phone{+16507230115}
\author{Marina Radulaski}
\affiliation{Spilker Center for Engineering and Applied Sciences, Stanford
University, Stanford CA 94305}
\author{Jan Petykiewicz}
\affiliation{Spilker Center for Engineering and Applied Sciences, Stanford
University, Stanford CA 94305}
\author{Konstantinos G. Lagoudakis}
\affiliation{Spilker Center for Engineering and Applied Sciences, Stanford
University, Stanford CA 94305}
\author{Ju-Hyung Kang}
\alsoaffiliation{Department of Materials Science and Engineering, Stanford
University, Stanford CA 94305}
\author{Mark Brongersma}
\alsoaffiliation{Department of Materials Science and Engineering, Stanford
University, Stanford CA 94305}
\author{Klaus Biermann}
\alsoaffiliation{Paul-Drude-Institut f\"{u}r Festk\"{o}rperelektronik,
Hausvogteiplatz 5-7 D-10117, Berlin, Germany}
\author{Jelena Vu\v{c}kovi\'{c}}
\affiliation{Spilker Center for Engineering and Applied Sciences, Stanford
University, Stanford CA 94305}

\keywords{nanophotonics, semiconductor microcavities, photonic crystals,
nonlinear optics}

\begin{document}

\begin{abstract}
We demonstrate second harmonic generation in photonic crystal cavities in
(001) and (111)B oriented GaAs.  The fundamental resonance is at 1800 nm,
leading to second harmonic below the GaAs bandgap. Below-bandgap operation
minimizes absorption of the second harmonic and two photon absorption of
the pump. Photonic crystal cavities were fabricated in both orientations
at various in-plane rotations of the GaAs substrate.  The rotation
dependence and farfield patterns of the second harmonic match simulation.
We observe similar maximum efficiencies of 1.2 \%/W in (001) and (111)B
oriented GaAs.\end{abstract}

%%%%%%%%%%%%%%%%%%%%%%% References %%%%%%%%%%%%%%%%%%%%%%%%%

%%%%%%%%%%%%%%%%%%%%%%%%%%  body  %%%%%%%%%%%%%%%%%%%%%%%%%%
\section*{}

Photonic crystal cavities are excellent candidates for nonlinear optical
devices, due to their low mode-volume and high quality (Q) factor. As
discussed previously \cite{liscidini_highly_2004,rodriguez_chi2_2007}, in
such optical cavities, the optical mode volume is small compared to the
material nonlinear coherence length, and the phase matching condition is
replaced by the requirement of large mode overlap between the relevant
optical modes.  This offers an additional advantage for III-V semiconductor
materials, which possess high nonlinearity but no birefringence. Many
different approaches have been taken in order to overcome this limitation
and take advantage of III-V materials, including growth of multiple quantum
wells to induce birefringence \cite{fiore_phase_1998}, orientation
patterning for quasiphasematching \cite{eyres_all-epitaxial_2001}, inversion
phase-matching in microdisks \cite{kuo_second-harmonic_2014}, form
birefringence in waveguides \cite{stievater_mid-infrared_2014}, surface
emitting cavities \cite{vakhshoori_blue-green_1991} and integration with
optical microcavities \cite{buckley_second_2013}. Integration of III-V
materials with photonic crystal cavities requires only standard
semiconductor processing \cite{khankhoje_modelling_2010}, while the geometry
of these cavities also allows easy integration of active gain media such as
quantum dots or quantum wells
\cite{rivoire_fast_2011,buckley_quasiresonant_2012,ota_nanocavity-based_2013},
as well as potential on-chip integration with detectors, switches and
modulators.

Experimentally, there have been many recent demonstrations of high
efficiency, low power $\chi^{(2)}$ nonlinear processes in resonant
microcavities, in particular second harmonic generation in microdisks
\cite{kuo_second-harmonic_2014,xiong_integrated_2011} and microrings
\cite{levy_harmonic_2011, pernice_second_2012} in materials such as GaAs,
GaN and AlN, as well as second harmonic generation and sum frequency
generation in photonic crystal cavities in materials such as InP
\cite{mccutcheon_experimental_2007}, GaP \cite{rivoire_second_2009}, GaAs
\cite{rivoire_fast_2011, ota_nanocavity-based_2013, buckley_second_2013} and
LiNbO$_3$ \cite{diziain_second_2013}. Millimeter sized lithium niobate
microdisks have also been used for high efficiency second harmonic
generation and ultra-low threshold optical parametric oscillators (OPOs)
\cite{ilchenko_low-threshold_2003,furst_naturally_2010,fortsch_versatile_2013}.

In order to achieve efficient nonlinear frequency conversion, it is
necessary to choose a material with a nonlinear susceptibility tensor
symmetry that matches the symmetry of the cavity modes well
\cite{vakhshoori_blue-green_1991, buckley_second_2013} (e.g. by choosing
crystal orientation).  For photonic crystal cavities, modes can be described
as having either TE-like or TM-like polarization
\cite{joannopoulos_photonic_2011}.  In the (001) orientation, the only
allowed polarization conversion is from TE-like to TM-like modes (or from
TM-like to TE-like). Moving to a different crystal orientation, such as the
(111) orientation, is equivalent to rotating the electronic crystal axes
relative to the photonic crystal axes, and therefore conversion from TE-like
to TE-like and TM-like to TM-like modes are also allowed. Additionally, it
is important to choose a transparency window that overlaps well with the
experimental frequencies. GaAs has a transparency window from around 900 nm
to 16 $\mu$m, and so is particularly useful for nonlinear frequency
conversion if all three wavelengths are within this range. Within this
frequency range, GaAs is preferable to wider bandgap semiconductors such as
GaP, as it has a stronger nonlinearity \cite{shoji_absolute_1997}, is easier
to grow in the (111) crystal orientation, and is compatible with bright gain
media such as InGaAs quantum wells, and efficient quantum emitters, such as
InAs quantum dots \cite{rivoire_fast_2011,ota_nanocavity-based_2013,
ota_self-frequency_2013}.

Here, we fabricate perturbed three hole defect (L3) photonic crystal
cavities in (001) and (111)B oriented GaAs. The fundamental mode is at
around 1800 nm, and thus the generated second harmonic is below the bandgap
of the GaAs, leading to minimal absorption \cite{palik_handbook_1985} and
two photon absorption, which was present in the previous studies in this
material \cite{buckley_second_2013, ota_nanocavity-based_2013} due to their
operation above the bandgap. The lack of absorption and other non-linear
absorption effects allows us to more easily simulate the second harmonic
mode, and to compare the simulations quantitatively with the experimentally
measured far-field momentum space (k-space) of the second harmonic emission.
While k-space measurements were performed in previous studies
\cite{mccutcheon_experimental_2007,rivoire_second_2009} in other materials,
here we expand upon the measurement and simulations of the generated modes,
matching the simulated and experimental results, and demonstrate that the
modes vary significantly with the photonic crystal cavity parameters and
effective nonlinear susceptibility tensor symmetry (i.e. GaAs crystal
orientation).  Therefore, the semiconductor crystal orientation can be
employed in addition to optical cavity design to improve the efficiency of
frequency conversion \cite{buckley_second_2013}.

\section*{Linear and nonlinear characterization of structures}
\label{section:exp}

\begin{figure}
\includegraphics[width = 12cm]{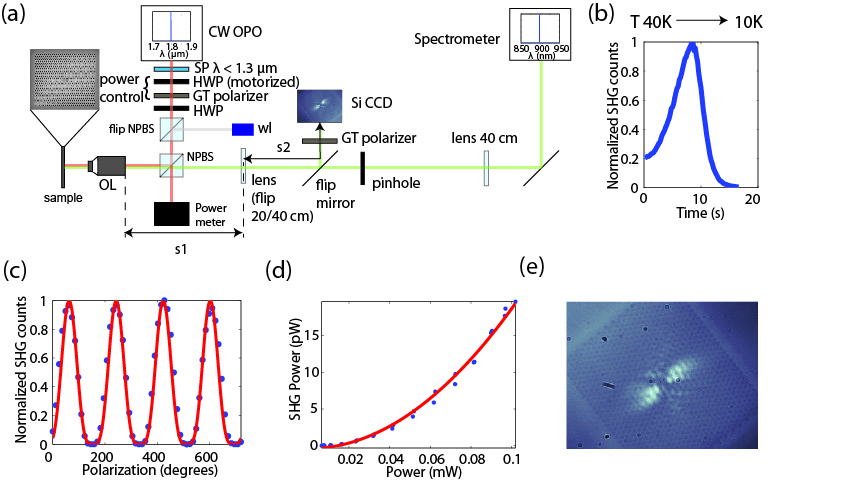}
\centering
{\caption{(a) Setup for generation of second harmonic and measurement of the k-space.
(NP)BS = (non-polarizing) beamsplitter, HWP = half wave plate, GT = Glan-Thompson,
SP = short pass filter, WL = white light source, OL = objective lens (b) Scanning the
 temperature of a photonic crystal structure in (111)B oriented GaAs between 40 and 10 K
 with the laser wavelength fixed at 1785 nm.  SHG signal was measured every 0.2 s
 (temperature change was not necessarily linear in time) and plotted versus time.
 This demonstrates the presence of a resonance in this wavelength range, although
 we can not extract the Q.
 (c) Scanning the input polarization to the cavity for the same cavity and wavelength as in part (b).
 (d) SHG power collected through the lens vs input power transmitted through the lens for a  structure in (001) oriented GaAs, the three hole defect at
 60$^\circ$ relative to the [110] or [1$\overline{1}$0] direction with resonant wavelength at 1799 nm.
 (e) Real space white light image of a photonic crystal structure in (001) oriented
 GaAs with the three hole defect at 45$^\circ$ to the [110] or [1$\overline{1}$0] axis with second harmonic emission also visible
 on Si CCD (10 mW pump power at 1750 nm).  The structure is about 25 $\times$ 20 $\mu$m.}
\label{fig:SHG_characterization}}
\end{figure}

Perturbed L3 photonic crystal cavities, as described in refs.
\cite{toishi_high-brightness_2009} and \cite{portalupi_planar_2010},
were fabricated in 165 nm thick (001) and (111)B =
($\overline{1}\overline{1}\overline{1}$) oriented GaAs membranes grown on an
AlGaAs sacrificial layer (0.878 \textbf{$\mu$}m thick in the (001) sample,
0.8 \textbf{$\mu$}m thick in the (111)B sample) on n-type doped substrates.
The cavities were fabricated using e-beam lithography and dry etching to
define the pattern, followed by HF wet etching to remove the sacrificial
layer as described previously \cite{khankhoje_modelling_2010}. The
fabricated structures had lattice constant $a$ = 560-620 nm  (resonant
wavelengths between 1730 nm and 1900 nm) and designed hole radius $r_1/a$ =
0.3, with perturbed hole radius $r_2/a$ = 0.33. We choose these parameters
in order to obtain structures with the second harmonic below the bandgap of
GaAs, but detectable on a Si CCD.  Fabricated photonic crystal cavities were
all characterized experimentally at the fundamental (1$^\mathrm{st}$
harmonic) wavelength with a broadband LED light source using a cross
polarized reflectivity method \cite{altug_photonic_2005}, with Q factors of
3000 to 4000 measured, with a simulated mode volume of V = 0.76
$(\lambda/n^3)$.

The setup for characterizing generated second harmonic is shown in Fig.
\ref{fig:SHG_characterization} (a), with the flip mirror down and the 40 cm
lens in place for a confocal measurement. Light from a continuous wave
optical parametric oscillator (OPO) was tuned to the resonant wavelength
(measured previously via cross-polarized reflectivity) of a particular
cavity, and was reflected from a beam splitter and coupled to the cavity at
normal incidence through a high numerical aperture (NA) objective lens. The
laser power was monitored in real-time via the transmission port of the beam
splitter, and the system was calibrated to calculate the power transmitted
through the objective.  The laser power incident on the structure was
controlled with a half wave plate (HWP) on a motorized rotation stage,
followed by a Glan-Thompson polarizer. The input polarization on the cavity
was then controlled by a second HWP after the polarizer.  The second
harmonic was collected through the same objective lens, and transmitted
through the beam splitter, where it was collected on a CCD spectrometer. The
losses through the objective lens, beam splitter and other optics were
measured and the spectrometer calibrated at the wavelength of the second
harmonic to obtain the second harmonic power emitted into the numerical
aperture of the objective lens.  To verify the SHG is a cavity effect,
rather than a bulk GaAs effect, we scanned the resonant wavelength of the
cavity across the laser wavelength by scanning the temperature of the
sample. While the OPO could be set to the wavelength of our choice, the
tuning was not smooth and therefore to obtain a scan over the resonance it
was more convenient to vary the resonant wavelength with temperature then to
scan the laser. For this measurement, the sample was mounted in a liquid
helium cryostat and a Zeiss objective with NA of 0.75 was used to couple
light to the structure. The set temperature of the cryostat was reduced from
40 K to 10 K (this range was found to be sufficient to scan across the
resonance while the cryostat remained most stable in this range). As the
temperature dropped slowly, the second harmonic power was measured at equal
time intervals.  A resulting peak is shown in Fig.
\ref{fig:SHG_characterization} (b). We only measured the actual temperature
at the start and end points, while the rate of change of temperature versus
time as well as the change in resonant wavelength versus temperature would
be necessary in order to perform an accurate fit; our previous experiments
\cite{rivoire_second_2009} show that the Q factor follows a Lorentzian
squared with the same Q factor measured in the reflectivity measurement,
which in this case was 4000. However, the measurement indicates that the
second harmonic process is sensitive to the resonant wavelength of the
cavity.

For subsequent measurements no cryostat was present and an Olympus objective
lens with NA 0.95 was used.  The OPO was set to the measured wavelength of
the cavity. The polarization incident on the cavity was also scanned by
varying the HWP angle, as shown in Fig. \ref{fig:SHG_characterization} (c).
The SHG intensity follows a $cos^4\theta$ dependence on HWP angle $\theta$
as shown by the red line fit, as expected from the L3 cavity, which is
strongly polarized in the y-direction (perpendicular to the line defined by
the three hole defect of the cavity). There is a phase offset as the x axis
of the cavity was not aligned to 0$^\circ$ for the polarizer. The second
harmonic power dependence on input power can be recorded by varying the
laser power at a particular wavelength with the HWP/polarizer combination;
the resulting curve for a typical cavity is shown in part (d), with a
quadratic fit in red.

\section*{Farfield measurements}
\label{section:farfields}
To obtain information about the second harmonic
modes, we imaged the k-space of the second harmonic signal in (001) and
(111)B oriented wafers for in-plane cavity rotations of 0$^\circ$,
30$^\circ$, 45$^\circ$, 60$^\circ$, 75$^\circ$ and 90$^\circ$ relative to
the [110] or [$1\overline{1}0$] direction (information about which was which
was not available) of the (001) oriented wafer and the [$1\overline{1}0$]
direction of the (111) oriented wafer. The setup was in 2$f$-2$f$
configuration as shown in Fig. \ref{fig:SHG_characterization} (a) with the
flip mirror up and the 20 cm lens in place such that $s_1$ = $s_2$ = 2$f$ =
40 cm. The position of the camera was optimized such that a sharp image of
the back aperture of the objective lens was generated. By flipping up the 40
cm lens, we could also image the real space signal on the camera as shown in
Fig. \ref{fig:SHG_characterization} (e).

\begin{figure}
\includegraphics[width=15 cm]{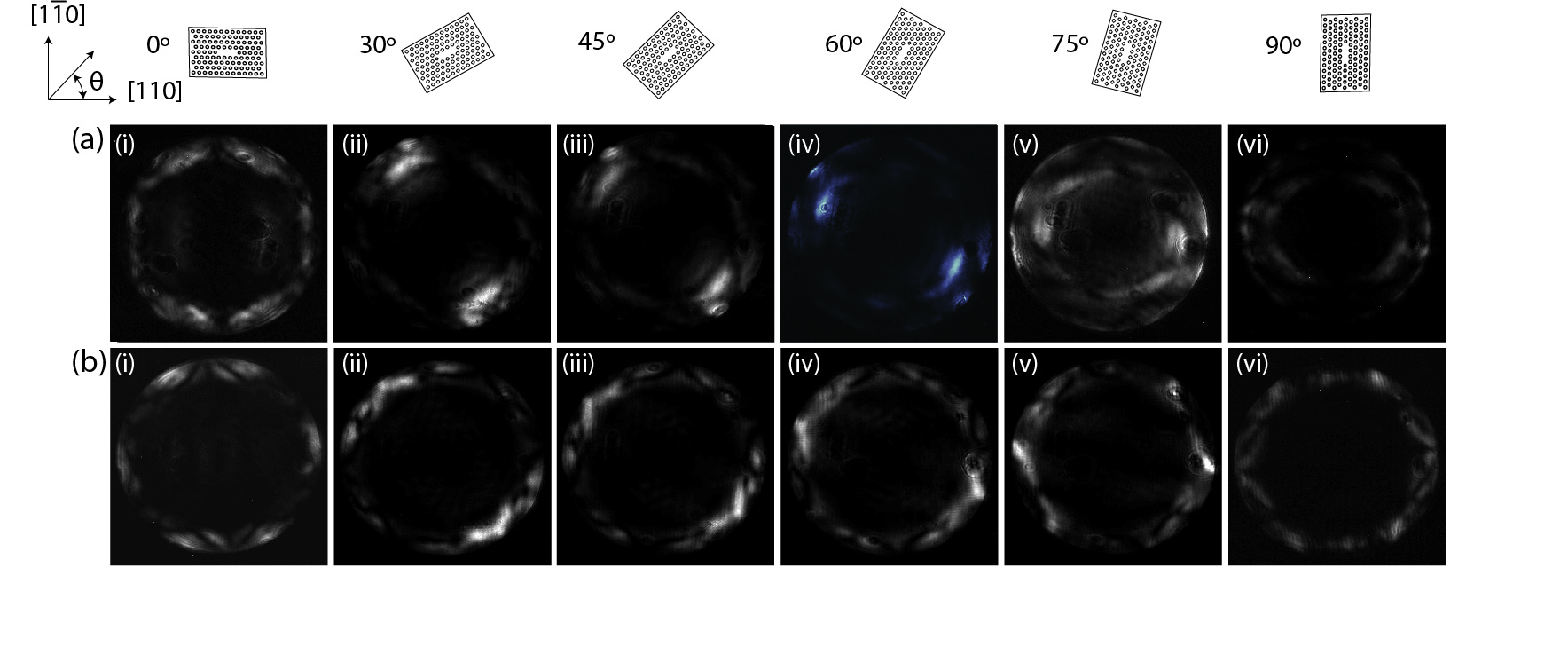}
\centering
{\caption{Experimental k-space profiles of second harmonic emission in (001) oriented GaAs L3 cavities with lattice constants
(a) 560 nm (mean fundamental resonance $\lambda_1$ = 1757.5 nm  and (b) 580 nm (mean $\lambda_1$ = 1800.3 nm) for in plane rotations of
(i) 0$^\circ$, (ii) 30$^\circ$, (iii) 45$^\circ$, (iv) 60$^\circ$, (v) 75$^\circ$, (vi) 90$^\circ$.}
\label{fig:100-farfield-rotations}}
\end{figure}

The measured farfields for lattice constant 560 nm (mean pump wavelength
$\lambda_1 = 1757.5$ nm, standard deviation 6.5 nm) and 580 nm (mean
$\lambda_1 = 1800.3$ nm, standard deviation 6.2 nm) in (001) oriented GaAs
are shown in Fig. \ref{fig:100-farfield-rotations} parts (a) and (b), with
rotations from 0$^\circ$ to 90$^\circ$ shown. Differences in intensity are
unrelated to the actual efficiency of the device; input power was adjusted
to keep the images below saturation of the camera, in order to help with
identification of the modes (see Supporting Information). The k-space images
for 0$^\circ$ and 45$^\circ$ in-plane rotations have very different spatial
patterns, which we expect since the overlap of the fundamental and second
harmonic mode changes with the rotation of the photonic crystal axes
relative to the crystal axes. However, this effect is complicated by slight
changes in other parameters. These additional changes for different
rotations are due to differences in fabrication at different angles relative
to the crystal axes and relative to the e-beam stage, which may cause
variations in hole shape with in-plane rotation angle. We observe that the
resonant wavelength within a particular rotation in the (001) orientation
varies by 3 nm, while the mean resonant wavelength decreases by 17 nm
between 0 and 75$^\circ$ rotations (with a slight increase again for
90$^\circ$). Simulations indicate that the membrane thickness and wavelength
have a large effect on the mode observed (see simulations section). This is
consistent with what we observe, as even the 3.5 \% change in lattice
constant from 560 nm to 580 nm with corresponding shift in resonant
wavelength from 1760 nm to 1800 nm has a noticeable effect on the farfield,
as can be seen in the difference between Fig.
\ref{fig:100-farfield-rotations} (a) and (b). Additionally, at these
wavelengths there is rapidly increasing absorption as the second harmonic
approaches the bandgap \cite{palik_handbook_1985}, which could also affect
the mode, in particular for the 560 nm lattice constant.

\begin{figure}
\includegraphics[width=15 cm]{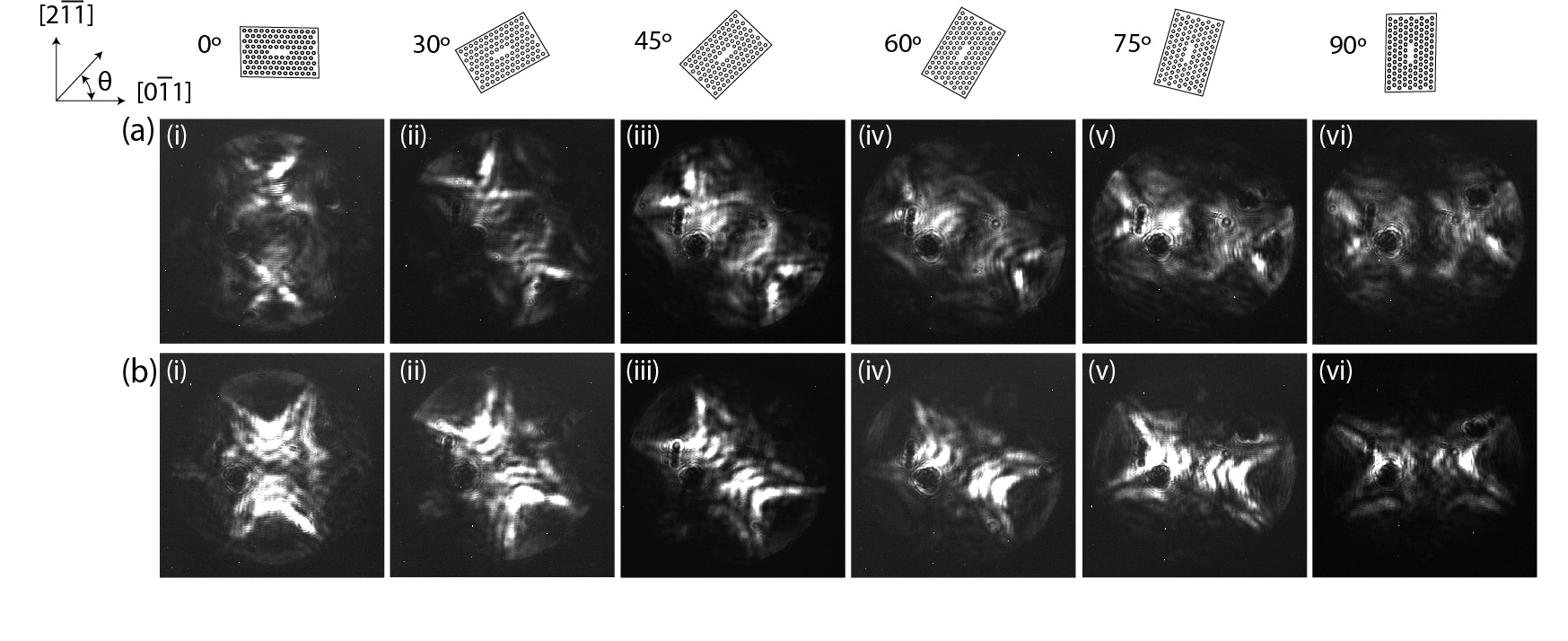}
\centering
{\caption{Experimental k-space profiles of second harmonic emission in (111)B oriented GaAs L3 cavities with lattice constants
(a) 600 nm (mean fundamental resonance $\lambda_1$ = 1769 nm  and (b) 620 nm (mean $\lambda_1$ = 1811 nm) for in plane rotations of
(i) 0$^\circ$, (ii) 30$^\circ$, (iii) 45$^\circ$, (iv) 60$^\circ$, (v) 75$^\circ$, (vi) 90$^\circ$.}
\label{fig:111-farfield-rotations}}
\end{figure}

Fig. \ref{fig:111-farfield-rotations} shows the same measurement for the
(111)B orientation, with (a) $a$ = 600 nm (mean $\lambda_1 = 1769$ nm,
standard deviation = 3 nm) and (b) $a$ = 620 nm (mean $\lambda_1 = 1811.1$
nm, standard deviation = 3.8 nm). Despite the fact that the membranes were
nominally the same thickness, in order to maintain the same resonant
wavelength as in the (001) orientation it was necessary to increase the
lattice constant by 40 nm, which indicates either a thinner membrane, larger
etched hole radius or larger refractive index; the exact cause was difficult
to determine via SEM images of the structures.  In this case, there is again
a decrease in average resonant wavelengths across rotations from 0 to
75$^\circ$, but in this case of only 7.5 nm, while the maximum variation
within a rotation is larger (6 nm, although on average it is lower than
this). The larger intra-rotation variation in the (111)B orientation is
likely due to a higher number of surface defects in this wafer.

The mode k-space distribution observed is very different for the two
orientations, and also changes less with rotation of the L3 in the plane of
the wafer in the (111)B-orientation, which matches our simulations (see
simulations section).

\section*{Second harmonic conversion efficiency}
\label{section:conversionefficiency}

\begin{figure}
\includegraphics[width=13 cm]{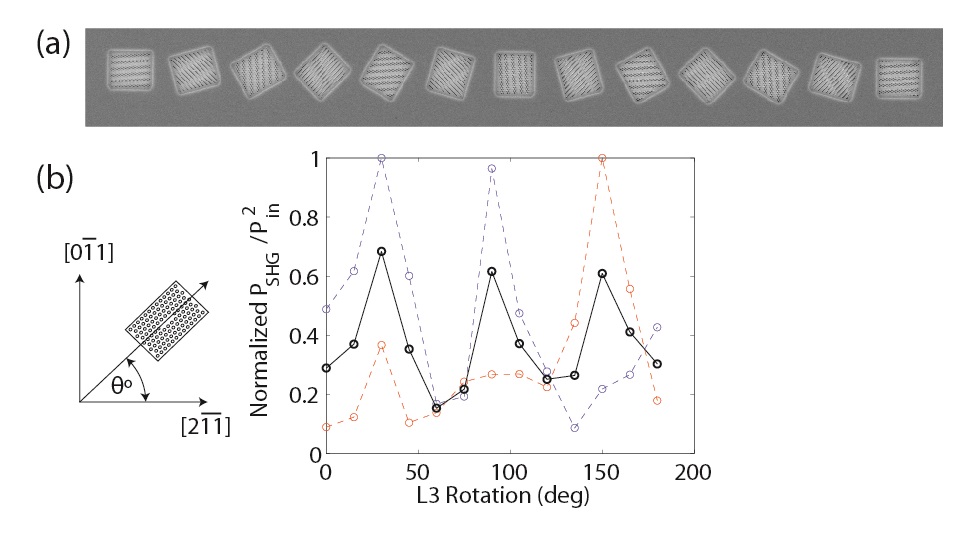}\centering
{\caption{(a) SEM of a single row of structures fabricated in (111)B GaAs, rotations every 15 degrees.
(b) Red and blue lines show normalized efficiency per unit power for two different rows as shown in (a),
resonant wavelength at 1800 nm.  Black line is the mean of the red and blue lines.}
\label{fig:cts_vs_rotation}}
\end{figure}

As second harmonic generation is a quadratic process at low powers, the
conversion efficiency per Watt ($P_{SHG}/P_{in}^2$) remains constant (see
Supporting Information). By measuring the second harmonic power versus input
power for a particular structure, we obtain a plot as shown in Fig.
\ref{fig:SHG_characterization} (d). Fitting this plot, we obtain a constant
value for $P_{SHG}/P_{in}^2$. We measured 12 cavities (including different
in-plane rotations) in each wafer orientation, $a$ = 580 nm in the (001)
oriented wafer, $a$ = 620 nm in the (111)B oriented wafer, as discussed in
the farfields section. The maximum measured conversion efficiency per Watt
for both (001) and (111)B oriented GaAs was 1.2 \%/W, although there was
again a large structure to structure variation even at a particular in-plane
rotation, perhaps due to the strong sensitivity to in- and out-coupling (see
Supporting Information), which will vary due to small variations in
structures as well as alignment. By comparison, previous studies in similar
structures in GaP \cite{rivoire_second_2009} reported a similar total
conversion efficiency per Watt of around 0.9 \%/W at telecommunications
wavelengths, and a much reduced efficiency per Watt of 0.002\%/W in GaAs at
telecommunications wavelengths \cite{buckley_second_2013}.  We discuss how
this compares to simulations in the simulations section.

Due to the symmetry of the nonlinear susceptibility tensor, the efficiency
of the second harmonic process will vary with in-plane rotation of the
photonic crystal cavity (see Supporting Information). To verify the rotation
dependence for structures in the (111)B orientation, we fabricated a second
chip with more cavities with in-plane rotations relative to the
$[2\overline{1}\overline{1}]$ direction.  Rotations were every 15$^\circ$
from 0$^\circ$ to 180$^\circ$, as shown in the SEM in Fig.
\ref{fig:cts_vs_rotation} (a) and lattice constant was 620 nm. The cavities
had mean fundamental resonant wavelength 1803.3 nm, with a standard
deviation of 1.7 nm. The stripes visible in the SEM are an artifact of the
SEM. We plot normalized efficiency per unit power versus rotation for two
rows of structures with the same parameters, shown in red and blue in Fig.
\ref{fig:cts_vs_rotation} (b). The plots were normalized to the maximum
value for each row, and the mean of the two is plotted in black, where we
see a $\pi/3$ periodicity as a function of in-plane rotation.  This is
expected due to to the fact that a $\pi/3$ rotation in the (111) orientation
is equivalent to an inversion of the crystal axes, combined with the $\pi$
symmetry of the photonic crystal cavity (see Supporting Information). Refs.
\cite{mccutcheon_experimental_2007} and \cite{rivoire_second_2009} have
previously reported variation in second harmonic signal in (001) oriented
III-V semiconductors depending on the in-plane rotation of the structure.

\section*{Simulations}
\label{section:simulations}

We next perform simulations in order to try to reproduce the farfield
k-space observed in experiment and the efficiency of the second harmonic
process.  This is done by first simulating the fundamental mode, and then
using this to generate a spatial polarization with which we can simulate the
second harmonic mode.  From these simulations we can estimate the in and
out-coupling efficiencies we should obtain using an objective lens with
numerical aperture of 0.95. Once we have the spatial profiles, coupling
efficiencies and Q factors of both modes, we can estimate the low power
efficiency of the device.  We additionally explore the parameter space
around our device in order to determine the sensitivity of the device to
design parameters, and to explore the possibility to engineer higher
efficiency devices.

The fundamental mode was simulated by finite difference time domain
simulations (FDTD).  As discussed in the farfields section, despite the fact
that we use nominally the same membrane thickness, to maintain the same
resonant wavelength we need to increase the lattice constant by 40 nm in
(111)B GaAs compared to (001) GaAs. In order to precisely match the
fundamental mode wavelengths to the simulation, the radius of the hole radii
and thickness of the membrane were adjusted around the designed values.
Simulations of the fundamental mode indicate that for a membrane thickness
of 165 nm, a radius of 0.28$a$ is consistent with the measured resonant
frequencies for (001) oriented GaAs structures, while a radius of 0.3$a$ is
consistent with the (111)B oriented GaAs resonant frequencies. The relative
size of perturbed holes was maintained constant. However, we found better
agreement with second harmonic simulations by varying the membrane
thickness.

The radiative ($Q_{rad}$) and total ($Q_{tot}$) Q factors of the photonic
crystal cavities were calculated from simulations, in order to obtain an
estimate for the cavity coupling efficiency. We take the coupling efficiency
$\eta_1 = f \cdot Q_{tot}/2Q_{rad}$, where $f$ is the fraction of the
radiated light (coupled to the cavity from vertically above it) through the
NA of the objective lens.  This fraction is calculated from the fraction of
the radiation vectors within the light cone that also have k-vectors within
the NA of the lens. We simulate the farfield by performing the fourier
transform the complex fields a distance $s$ above the surface of the slab as
described in references\cite{vuckovic_optimization_2002} and
\cite{taflove_computational_2005}.

In order to simulate the second harmonic mode, we generate a polarization
from the fundamental mode, which is the initial excitation for our
simulation:
\begin{equation}
P_i = \epsilon_{bin}\chi^{(2)}_{ijk} E_j E_k\\
i = x,y,z
\end{equation}
where $\epsilon_{bin}$ is 1 wherever there is semiconductor, and 0 in air.
This generated polarization will be different in the case of the (001) and
(111) orientations and will also depend on the rotation of the L3 cavity
with respect to the crystal axes in the plane of the wafer (see Supporting
Information).  We calculate the farfield for the simulated modes to compare
these to experiment. These modes were particularly sensitive to the membrane
thickness of the structure. Therefore we varied the membrane thickness of
the cavity around the experimental values while maintaining the correct
fundamental wavelength. For each simulation, we calculated the overlap of
simulated and experimental k-space in order to find the best match (see
Supporting Information).

Implicit so far has been the assumption that the nonlinearity is due to bulk
effects, while the nanocavity employed has is a significant surface to
volume ratio.  The symmetry of the surface is different than the bulk
$\chi^{(2)}$ effect \cite{stehlin_optical_1988}, although in the case of
(001)-GaAs the (001) surface (which is the main surface seen by the mode)
nonlinearity still prevents TE-TE mode coupling.  Our simulations of surface
versus bulk overlap indicate that while the surface may contribute slightly
to the second harmonic signal, the main contribution is from the bulk
$\chi^{(2)}$ nonlinearity. This is in agreement to the conclusions found in
previous studies of photonic crystal cavities
\cite{mccutcheon_resonant_2005}. Experimentally, by performing measurements
on a number of structures rotated  in-plane , it should be possible to
distinguish the surface from bulk contributions \cite{stehlin_optical_1988}.
However, this effect is complicated due to the fact that the second harmonic
mode is very sensitive and will change depending on the excitation, and any
measurement would need to be statistical over many structures, as the
polarization is set by the in-plane rotation of the cavity.  As there is
significant structure to structure variation due to fabrication
imperfections, such an effect may be impossible to measure.    Even the
rotation dependence of the surface will depend significantly on the presence
of any native oxides or surface contamination \cite{armstrong_optical_1992}.

\begin{figure}
\includegraphics[width=15 cm]{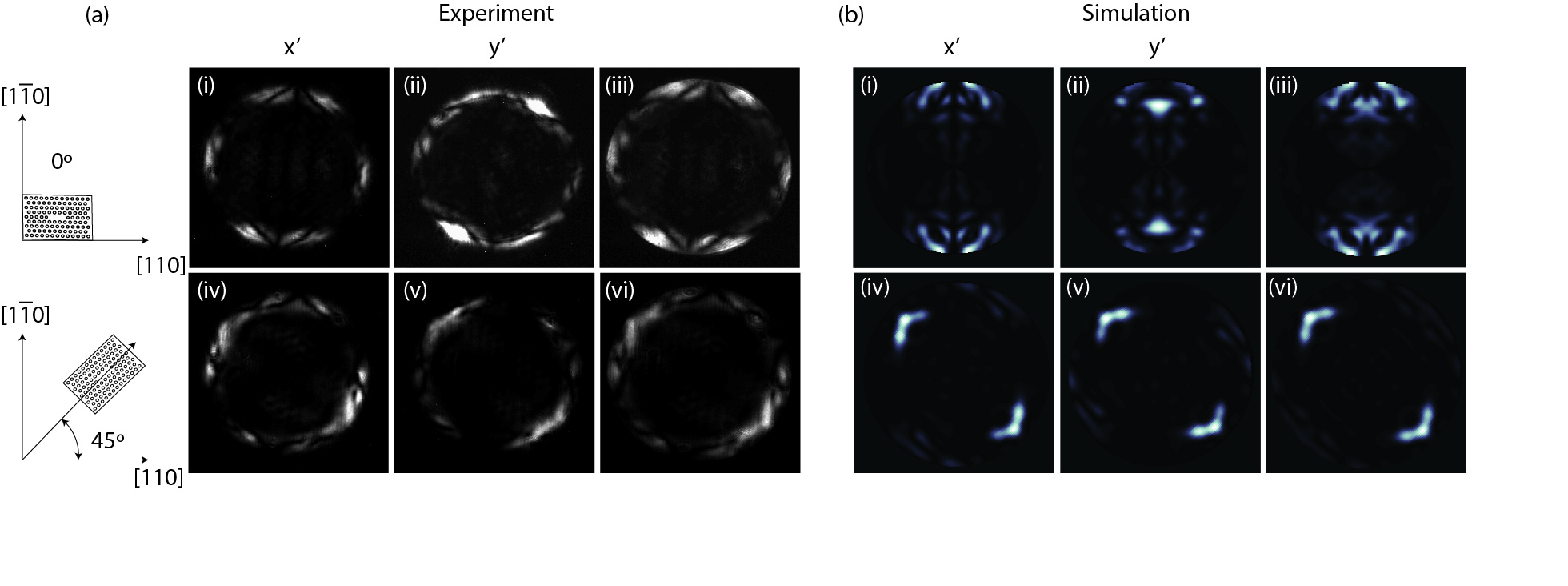}
{\caption{(a) Experimentally measured farfield of generated second harmonic on
(001) GaAs wafer, as a function of  L3 photonic crystal cavity in-plane rotation.
The top/bottom row corresponds to L3 cavity at 0/45$^\circ$ relative  to the
cleave ([110] or equivalent) axes. (i) and (iv) show the
farfield for just the x polarization, while (ii) and (v) show the farfield for
just the y polarization and (iii) and (vi) show the total image.  Part (b)
shows the simulated farfields for the parameters used in the experiment.}
\label{fig:100-0-90-all}}
\end{figure}

\begin{figure}
\includegraphics[width=15 cm]{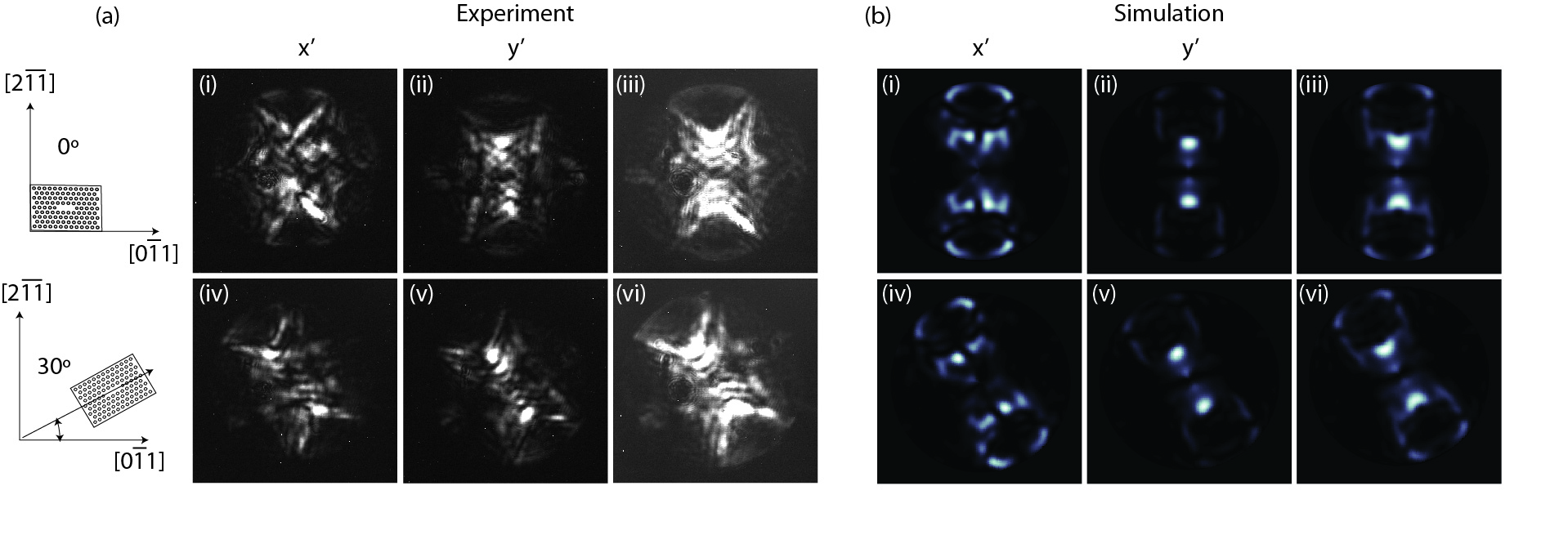}
{\caption{(a)  Experimentally measured farfield of generated second harmonic on
(111) GaAs wafer, as a function of  L3 photonic crystal cavity in-plane rotation.
The top/bottom row corresponds to L3 cavity at 0/30$^\circ$ relative  to the
cleave ([11$\overline{2}$] or equivalent) axes. (i) and (iv) show the
farfield for just the x polarization, while (ii) and (v) show the farfield for
just the y polarization and (iii) and (vi) show the total image.  Part (b)
shows the simulated farfields for the parameters used in the experiment.}
\label{fig:111-0-90-all}}
\end{figure}

Figs. \ref{fig:100-0-90-all} and \ref{fig:111-0-90-all} show (a)
experimental and (b) simulated k-space images for (001) and (111)
orientations, with simulations plotting up to NA = 0.95. For the (001)
orientation, we compare the experimentally measured and simulated k-space
images for both 0 and 45 degree rotations for the cavity mode with resonant
frequency at 1800 nm, while for (111) orientation we compare 0 and 30 degree
rotations with simulations.  The (001) oriented cavity simulation was
consistent with a membrane thickness of 165 nm.  The (111) orientation
simulation indicated a membrane thickness of 150 nm.

For triply resonant cavity processes at low input powers, the second
harmonic power $P_{SHG}$ is proportional to the input power $P_{in}$ as
\begin{equation}
P_{SHG} = \frac{32 \eta_1^2\eta_2Q_1^2Q_2|\beta_1|^2}{\omega_1}P_{in}^2
\label{eq:p_shg}
\end{equation}
(see Supporting Information) where $Q_1$ and $Q_2$ are the quality factors
of the fundamental and second harmonic modes, $\eta_1$ and $\eta_2$ are the
input/output coupling efficiencies of the fundamental and second harmonic
modes and $|\beta|^2$ is a non-linear overlap integral given by
\cite{rodriguez_chi2_2007, burgess_design_2009},
\begin{equation}
\beta_1 = \frac{1}{4} \frac{\iiint d^3 x \epsilon_0 \sum_{ijk}
\chi^{(2)}_{ijk} E_{1i}^{*}  \left(E^{*}_{1j} E_{2k} + E^{*}_{1k}E_{2j}\right)}
{\iiint d^3 x \epsilon |E_1|^2 \sqrt{\iiint d^3 x \epsilon |E_2|^2}}
\label{eq:beta}
\end{equation}

whose value for a particular set of modes depends on the magnitude and
symmetry of $\chi^{(2)}$.  From equation 2 we can see that the
conversion efficiency is very sensitive to the Q factors of both modes, as
well as to the coupling efficiency.

\begin{figure}
\includegraphics[width=15 cm]{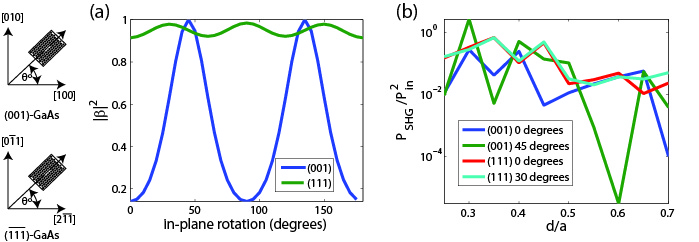}\centering
{\caption{(a) Normalized simulated nonlinear overlap between fundamental and
 SHG modes versus in-plane rotation for (111) and (001) wafer orientations
 relative to the [$2\overline{1}\overline{1}$] and [100] directions respectively.
(b) Simulated efficiency per unit power (W$^{-1}$) versus membrane thickness
for a particular L3 cavity design, and for (111) and (001) wafer orientations
with different in-plane cavity rotations relative to the cleave axis.}
\label{fig:beta_simulations}}
\end{figure}

Using these simulations, we can make estimates of this nonlinear overlap
integral, as well as of the total Q and coupling Q factors of the second
harmonic mode. We verify that the simulations give us the expected in-plane
rotation dependence in Fig. \ref{fig:beta_simulations} (a), where we plot
the nonlinear overlap $|\beta|^2$ (each normalized to its maximum value)
versus the in-plane rotation for (001) and (111) orientations, and obtained
the expected 60 degree and 90 degree symmetries. From equation
3, we can calculate the simulated efficiency per Watt input
power.

The fundamental and second harmonic modes were simulated for increasing
$d/a$ (in experiment corresponding to an increase in membrane thickness,
with resonant wavelength maintained constant).  The plot of
$P_{SHG}/P_{in}^2$, shown in Fig. \ref{fig:beta_simulations} (b), shows that
the geometry of the second harmonic mode is very important in calculating
the overall efficiencies. From this plot, we estimate efficiencies of the
order of 10 \%/W for both (001) and (111) orientations, compared to the
experimentally measured values of the order of 1\%/W. The internal
conversion efficiency can also be estimated using the simulated coupling
efficiency values (around 30$\%$) to give a simulated internal efficiency of
370$\%$/W.  These calculated coupling efficiencies are likely higher than
the experimentally achieved coupling efficiency, and could explain the
discrepancy between simulated and measured values.

From this plot, we see that the particular second harmonic mode excited is
very important in determining the efficiency of the device.  For example,
depending on the particular second harmonic mode, either the 0 or 45$^\circ$
in-plane rotation in the (001) orientation can have higher efficiency.  The
(111) and (001) orientations also vary in relative efficiencies depending on
parameters, and therefore with knowledge of the second harmonic mode we can
use the wafer orientation in order to engineer higher efficiency devices.
 Having good control over these modes would also allow us to use
microcavities in order to measure nonlinear properties of new materials,
without needing to develop phase-matching and quasi-phasematching
techniques. However, this is challenging in these geometries as the
particular mode is very sensitive to the parameters of the device.

\section*{Engineering higher efficiency devices}
\label{section:engineering}

The experimentally achieved efficiency of second harmonic generation in
photonic crystal cavities has been limited by the difficulty of engineering
multiple high quality factor modes with a high degree of overlap
\cite{mccutcheon_experimental_2007,
rivoire_second_2009,diziain_second_2013}. This difficulty arises because the
bandgap of photonic crystals does not span a sufficiently large frequency
range for $\chi^{(2)}$ processes, and there are no significant higher order
bandgaps. This means that only one of the modes of the photonic crystal
cavity in the process is well defined and has high Q. As discussed in this
work and others
\cite{mccutcheon_experimental_2007,rivoire_second_2009,buckley_quasiresonant_2012,diziain_second_2013},
the second harmonic couples to leaky air band modes perturbed by the
presence of the cavity which have low Q factors, low overlap with the
fundamental mode, and are difficult to couple to for processes such as
difference frequency conversion and optical parameteric oscillation. The SHG
process goes as the square of the fundamental, so as shown in Eq. 2 it is
important that the fundamental mode have a large Q.  However, since we are
coupling at normal incidence, increasing the Q factor will in general lead
to a decrease in the coupling efficiency, and therefore increasing the Q
significantly beyond a few thousand does not lead to further improvement.
Creating a well defined coupling channel, such as a waveguide coupled to the
photonic crystal cavity, would allow us to obtain increased benefit from
higher Q.  For the current structure, a better understanding of these modes
outside the bandgap in photonic crystals may help to engineer higher
efficiency frequency conversion. The additional degree of freedom of
choosing the symmetry of the effective $\chi^{(2)}$ is also beneficial in
the design of resonators. Where individual resonators with high overlap and
Q factors may not be sufficient, the ability to engineer doubly resonant
microcavities opens up the possibility for generating highly nonlinear
materials that phasematch via engineered dispersion
\cite{xu_propagation_2000} or quasi-phasematch
\cite{dumeige_quasi-phase-matching_2011} using coupled resonator arrays.
There have been several proposals for designing photonic crystal cavities
with multiple high Q resonances and large frequency separations
\cite{zhang_ultra-high-Q_2009,burgess_design_2009,thon_polychromatic_2010,mccutcheon_broadband_2009,
rivoire_multiply_2011, rivoire_multiply_2011-1}; such a cavity could improve
the efficiency of the process by several orders of magnitude.

Integration of nonlinear microcavities with quantum dots for quantum optics
and quantum information processing is also of current interest. Cavities
such as those used in this work could be useful for high signal to noise
resonant excitation of InAs QDs \cite{rivoire_fast_2011}. Additionally,
frequency conversion of flying qubits to telecommunications wavelength or to
optimal detection wavelengths is desirable \cite{buckley_engineered_2012},
and has been demonstrated using off-chip periodically poled lithium niobate
waveguides. Self-frequency conversion of high density quantum dots in a
photonic crystal cavity has been recently demonstrated,
\cite{ota_nanocavity-based_2013, ota_self-frequency_2013}, but
demonstrations of frequency conversion of single quantum dots coupled to
on-chip microcavities has yet to be demonstrated.  In particular, frequency
conversion between InAs QD wavelengths and telecommunications wavelength
requires intra-cavity difference frequency generation, which is again
challenging, due to the long wavelength of the pump, and difficulties in
engineering photonic crystal cavities with well defined and overlapping
modes at sufficiently large wavelength separations. This work demonstrates
the characterization and operation of these cavities in GaAs at longer
wavelengths than previously demonstrated.

Creating a highly nonlinear element such as a photonic crystal in a
$\chi^{(2)}$ material is also of interest itself for quantum information
processing \cite{mabuchi_qubit_2012}, for generation of single photons via
photon blockade \cite{majumdar_single-photon_2013} or for strongly coupling
photons at two different wavelengths \cite{irvine_strong_2006}.

\section*{Conclusion}
We demonstrate second harmonic generation below the bandgap in photonic
crystal cavities in (001) and (111)B oriented GaAs. We fabricate photonic
crystal structures in both (111)B and (001) oriented GaAs at different
orientations with respect to the crystal axes of the GaAs substrate, and
match the rotation dependence and farfield patterns to simulation. We
discuss how these results are relevant to engineering higher efficiency
on-chip nonlinear frequency conversion in photonic crystal cavities.

\section*{Acknowledgements}

Financial support was provided by the Air Force Office of Scientiﬁc
Research, MURI Center for multi-functional light–matter interfaces based on
atoms and solids, National Science Graduate Fellowships, and Stanford
Graduate Fellowships. This work was performed in part at the Stanford
Nanofabrication Facility of NNIN supported by the National Science
Foundation under Grant No. ECS-9731293, and at the Stanford Nano Center.
J.V. also thanks the Alexander von Humboldt Foundation for support. K.G.L.
acknowledges support from the Swiss National Science Foundation.

\appendix

\section*{Appendix}

\section{Photonic crystal polarization and semiconductor orientation}
\label{app:100_vs_111}

The vertical symmetry of the photonic crystal cavity forces all modes to be
either primarily transverse electric (TE)-like or transverse magnetic
(TM)-like \cite{joannopoulos_photonic_2011}. For a TE-like mode, $E_x$,
$E_y$, and $H_z$ obey even symmetry in the vertical direction about the
center of the slab while $E_z=H_x=H_y=0$ at the central $xy$ plane and are
anti-symmetric about this plane (where we define $z$ as the direction normal
to the wafer, and $x$ and $y$ to be in the plane of the wafer). For TM-like
modes, $E_z$, $H_x$, and $H_y$ obey even symmetry in the vertical direction
about the center of the slab while $E_x=E_y=H_z=0$ at the central $xy$ plane
and are anti-symmetric about this plane.

For (100) oriented III-V semiconductors, the second order nonlinear
susceptibility tensor $d_{eff}$ is given by
\begin{equation}
d_{eff} =
\left( \begin{array}{cccccc}
0 & 0 & 0 & d_{41} & 0 & 0 \\
0 & 0 & 0 & 0 & d_{41} & 0 \\
0 & 0 & 0 & 0 & 0 & d_{41} \end{array} \right)
\end{equation}
where the generated polarization at a sum frequency can be calculated by
\begin{equation}
\left( \begin{array}{c}
P_x(2 \omega) \\
P_y(2 \omega) \\
P_z(2 \omega) \\ \end{array} \right) = 2 \varepsilon_0 d_{eff}\left( \begin{array} {c}
E_x(\omega)^2 \\
E_y(\omega)^2 \\
E_z(\omega)^2 \\
2E_y(\omega)\cdot E_z(\omega) \\
2E_x(\omega)\cdot E_z(\omega) \\
2E_y(\omega)\cdot E_x(\omega) \\ \end{array} \right)
\label{eq:d_100}
\end{equation}

For GaAs the $d_{eff}$ matrix found in the literature is defined such that
$x, y, z$ are the (100), (010) and (001) axes of the crystal structure
(while the cleave axes in GaAs are the (110) and equivalent planes).  This
means that for the case of (001) GaAs the $x$ and $y$ coordinates are in the
same plane as two of the major crystal axes, and can be chosen to be aligned
with the crystal axes.  Examining the mode overlap integral reveals that for
the process of second harmonic generation, a TE-like mode may only couple to
a TM-like mode if the wafer is normal to the [100], [010], or [001]
(equivalent) directions, as in standard (001) oriented wafers.

In the case of (111) GaAs, the plane of the wafer is no longer the same as
the plane of crystal axes.  The values of E-field can be either transformed
to this coordinate system or a new $d_{eff}$ matrix can be derived with
$x'$, $y'$ in the plane of the wafer.  Rotating from the (001) to the (111)
plane can be done by applying the following steps: (1) rotation about the
z-axis of 45 degrees (2) rotation through an angle of
$\arccos{\left(\frac{1}{\sqrt{3}}\right)}$ about the y-axis.

This gives the $d_{eff}$ matrix
\begin{equation}
\label{eq:d_111}
\left( \begin{array}{c}
P'_x(2\omega) \\
P'_y(2\omega) \\
P'_z(2\omega) \\ \end{array} \right) =
2 \varepsilon_0 d_{eff,111}\left( \begin{array} {c}
E_x(\omega)^{\prime2} \\
E_y(\omega)^{\prime2} \\
E_z(\omega)^{\prime2} \\
2E'_y(\omega)\cdot E'_z(\omega) \\
2E'_x(\omega)\cdot E'_z(\omega) \\
2E'_y(\omega)\cdot E'_x(\omega) \\ \end{array} \right)
\end{equation}
The calculated $d_{eff,111}$ is given by
\begin{equation}
d_{eff} = \left( \begin{array}{cccccc}
-\frac{1}{\sqrt{6}} & \frac{1}{\sqrt{6}} & 0 & 0 & -\frac{1}{\sqrt{3}} & 0 \\
0 & 0 & 0 & -\frac{1}{\sqrt{3}} & 0 & -\frac{1}{\sqrt{6}} \\
-\frac{1}{2\sqrt{3}} & -\frac{1}{2\sqrt{3}} & \frac{1}{\sqrt{3}}& 0 & 0 &
0 \end{array} \right)\cdot d_{41}
\end{equation}
We can see that this new $d_{eff}$ matrix leads to coupling between TE-like
polarizations, e.g. $P'_y = -\frac{1}{2\sqrt{3}} d_{41} E'_y E'_z
-\frac{1}{\sqrt{6}} d_{41} E'_y E'_x$. The tensor leads to the expected 120
degree symmetry, with the effective $x$ and $y$ axes along the
[11$\overline{2}$] and [$\overline{1}$10] directions.

\section{Calculation of low power efficiency}
\label{app:calculations}

 Following refs. \cite{rodriguez_chi2_2007} and
\cite{burgess_difference-frequency_2009}, the coupled mode equations for a
$\chi^{(2)}$ nonlinearity with two modes, one at fundamental frequency
$\omega_1$ and the second at the second harmonic frequency $\omega_2$, are
\begin{equation}
\frac{dA_1}{dt} = - \frac{\omega_1}{2Q_1} A_1 + i \omega_1 \beta_1 A_2 A^{*}_1 + \sqrt{\frac{\eta_1\omega_1}{Q_1}}S_{1+}
\end{equation}

\begin{equation}
\frac{dA_2}{dt} = - \frac{\omega_2}{2Q_2} A_2 + i \omega_2 \beta_2 A_1^2
\end{equation}

where $A_k$ is the time independent complex amplitude of the $k$the mode,
with $|A_k|^2$ normalized to electromagnetic energy stored in the mode.
$\beta_1 = \beta_2^*$ is the nonlinear coupling between the two modes, and
is given by the nonlinear overlap integral in Eq. 3 in the text
\cite{rodriguez_chi2_2007,burgess_difference-frequency_2009}. $S_{i\pm}$ is
the amplitude of the incoming ($+$) or outgoing ($-$) wave, and in this case
$S_{2+} = 0$.

\begin{equation}
S_{2+} = -S_{2-} + \sqrt{\frac{\eta_2\omega_2}{Q_2}}A_2\\
\rightarrow |S_{2-}|^2 = \frac{\eta_2\omega_2}{Q_2}|A_2|^2
\end{equation}

In steady state, the second harmonic output power in terms of $|A_1|$
\begin{equation}
|S_{2-}|^2 = 2 \eta_2 Q_2 \omega_1 |\beta_1|^2 |A_1|^4
\end{equation}
and the input power in terms of $|A_1|$
\begin{equation}
|S_{1+}|^2 = \frac{Q_1}{\eta_1 \omega_1} \left(\frac{\omega_1}{2 Q_1} + \omega_1 Q_2 |\beta_1|^2  |A_1|^2\right)^2 |A_1|^2
\end{equation}
Therefore, at low input powers, when $|A_1|^2<<\frac{1}{2Q_1 Q_2|\beta|^2}$
we have that
\begin{equation}
P_{in,lp} = \frac{\omega_1}{4\eta_1Q_1}|A_1|^2
\end{equation}
and therefore $P_{SHG}$ vs. $P_{in}$ is quadratic at low powers, with
efficiency per unit input power a constant given by
\begin{equation}
\frac{P_{SHG}}{P_{in,lp}^2} =  \frac{32 \eta_1^2\eta_2Q_1^2Q_2}{\omega_1}|\beta_1|^2.
\end{equation}

\section{Simulation inputs}
\label{app:simulation_inputs}

\begin{figure}[T]
\includegraphics[width = 10cm]{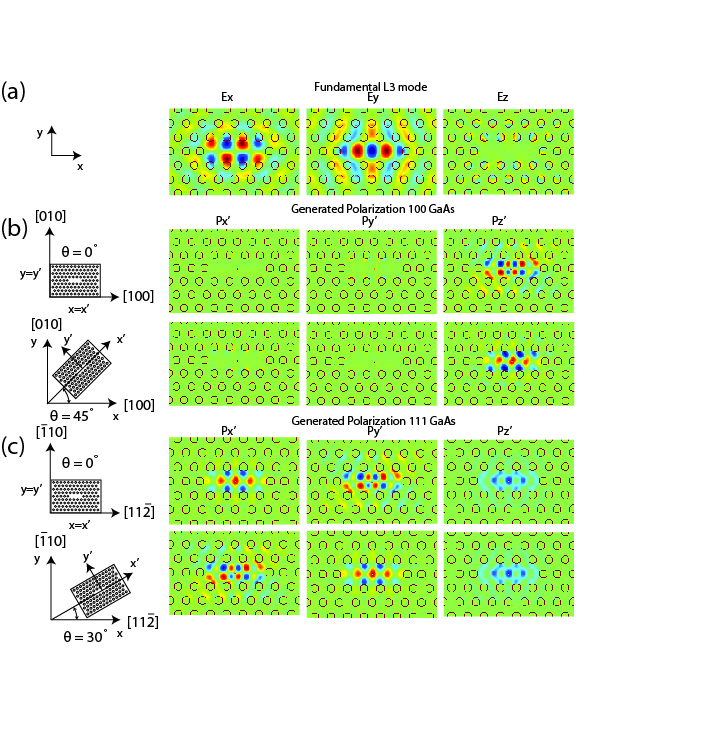}
{\caption{(a) Fundamental mode electric field profile at the vertical (z) center of the photonic crystal
(b) Central slice of input polarization for (001) GaAs simulations for 0 and 45 degree in-plane rotations.
(c) Central slice of the input polarization for (111) GaAs simulations for 0 and 30 degree in-plane rotations.}
\label{fig:L3_fields_polarizations}}
\end{figure}

Fig. \ref{fig:L3_fields_polarizations} shows the generated polarizations at
the center of the photonic crystal membrane for the (001) orientation for
0$^\circ$ and 45$^\circ$ in plane rotations and in the (111) orientation for
0$^\circ$ and 30$^\circ$ in-plane rotations. We generate this field profile
for the full 3D space of the simulation using a current sources, and allow
the field to evolve in time. The refractive index is changed in this
simulation to match the refractive index at the second harmonic wavelength
\cite{palik_handbook_1985}.

\section{Far field simulation and experimental image comparison}
\label{app:farfield_comparison}

\begin{figure}[htbp]
  \centering
  \includegraphics[width=10cm]{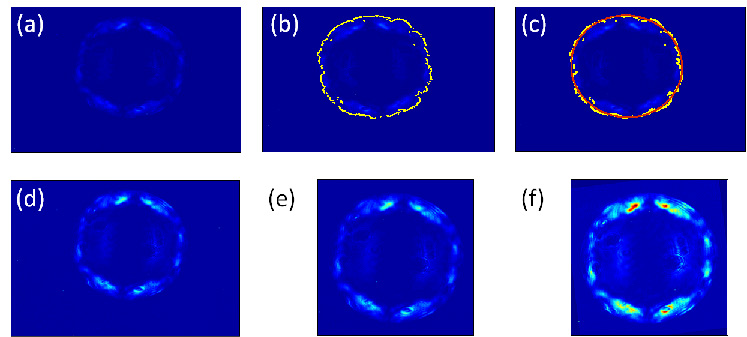}
\caption{Processing of an experimental image. (a) Original experimental image. (b) Identified edge contour of the signal. (c) Ellipse fit to the contour. (d) Resized image (a). (e) Cropped image (d). (f) Rotated image (e). }
\label{fig:paper1}
\end{figure}

\begin{figure}[htbp]
  \centering
  \includegraphics[width=14cm]{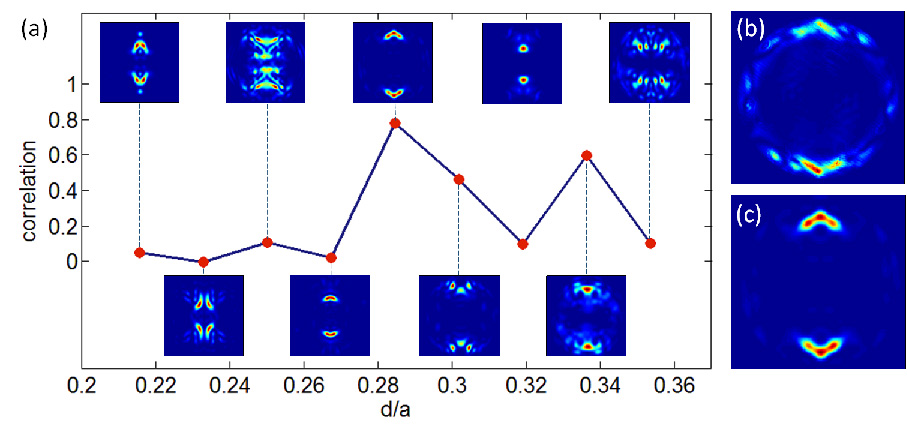}
\caption{a) Correlation between a sample processed experimental image and simulation images obtained for varying thickness to lattice constant ratio. (b) Processed experimental image from the comparison in (a). (c) Simulation image for $d/a = 0.284$ that maximizes the correlation in (a) with value $c = 0.81$. }
\label{fig:da_analysis}
\end{figure}

 To obtain theoretical understanding of the
SHG process in the cavities, we analyze far field image of the second
harmonic mode, by comparing experimental and simulation data.

Due to alignment imperfections of beam to the camera, and the photonic
crystal relative to the x-y axes, and deflection of the beam by the
polarizer, the experimental images are skewed and rotated (Fig.
\ref{fig:paper1} (a)) compared to the coordinate system used in the
simulation. To facilitate a successful image comparison, we first develop a
data-specific algorithm to resize and rotate the experimental image. The
data-specific content in this case is the k-space of the second harmonic
emission, the outer edge of which is defined by the edge of the image of the
back aperture of the objective lens. Outside of this aperture there is
negligible intensity. Due to imperfect alignment, the image of this aperture
is an ellipse instead of a circle, which varies from image to image due to
alignment adjustments that were made over the course of the experiment. We
identify this shape by finding the edge contour of the signal region, where
the signal value jumps (Fig. \ref{fig:paper1} (b)). Then, we fit that
contour to an ellipse (Fig. \ref{fig:paper1} (c)), and use the fit
parameters (ratio between long and short ellipse axes) to rescale the image
and obtain a circular signal (Fig. \ref{fig:paper1} (d)). We use the
coordinates of the center and radius of the new circle to crop the image
around the relevant signal (Fig. \ref{fig:paper1} (e)). Finally, we rotate
the image to match the coordinates of the simulation (Fig. \ref{fig:paper1}
(f)).

Comparison between the processed experimental and simulation images was done
by the following correlation algorithm
$$c = \frac{\sum_m \sum_n (A_{mn}-\bar A)(B_{mn}-\bar B)}{\sqrt{\sum_m \sum_n (A_{mn}-\bar A)^2(B_{mn}-\bar B)^2}}.$$
To find the best correlation between the two images, we vary zoom, rotation,
and x and y translation of the experimental image, recording the maximal
correlation value as a figure of merit. Fig. \ref{fig:da_analysis} (a) shows
the dependence of the correlation value between a processed experimental
image (Fig. \ref{fig:da_analysis}) ((001) orientation, $a$ = 580 nm,
in-plane rotation 45 degrees) and simulated patterns for increasing slab
thicknesses with resonant wavelength maintained constant (insets in the
plot), and are consistent with a simple visual inspection. Fig.
\ref{fig:da_analysis} (c)) shows the simulation pattern with highest
correlation value to the processed experimental image in Fig.
\ref{fig:da_analysis} (b).

\providecommand*\mcitethebibliography{\thebibliography}
\csname @ifundefined\endcsname{endmcitethebibliography}
  {\let\endmcitethebibliography\endthebibliography}{}

\end{document}